\documentclass[12pt]{article}
\usepackage{epsfig}

\newcommand{\beao}{\begin{eqnarray*}}
\newcommand{\eeao}{\end{eqnarray*}}
\newcommand{\be}{\begin{equation}}
\newcommand{\ee}{\end{equation}}
\newcommand{\bea}{\begin{eqnarray}}
\newcommand{\eea}{\end{eqnarray}}
\newcommand{\beq}{\begin{eqnarray}}
\newcommand{\eeq}{\end{eqnarray}}

\newcommand{\la}{\lambda}

\begin{document}

\title{The type of the phase transition \\ and coupling values in $\lambda~ \phi^4 $ model}
\author{M.~Bordag, V.~Demchik, A.~Gulov and V.~Skalozub}

\maketitle

\begin{abstract}
The temperature induced phase transition is investigated in the
one-compo\-nent scalar field $\phi^4$ model on the lattice.
Using the GPU cluster a huge amount of Monte Carlo simulation
data is collected for a wide interval of coupling values. This gives a possibility to
determine the  low bound on the coupling constant
$\la_0$ when the transition happens and investigate its type. We found that for the values
of $\la$ close to this bound a weak-first-order phase transition
takes place. It converts into a second order one with the
increase of $\la$. A comparison with the results obtained in analytic and
numeric calculations by other authors is given.
\end{abstract}






\section{Introduction}
Scalar field models with a spontaneous symmetry breaking (SSB) are considered in various
fields of physics, like quantum field theory, collective phenomena, quantum dots,
high-temperature superconductivity, etc. They often serve as toy models to develop both
the analytic calculation schemes and the numeric simulation techniques to describe a wide
class of temperature induced phase transitions. Multi-component scalar field with the
orthogonal symmetry $O(N)$ is a popular choice for these investigations. The simplest
model contains one-component scalar field and can be called the $O(1)$-model.

There is a long history of studying phase transitions in the $O(N)$ models (see
Refs.~\cite{ZinnJustin:1996cy}-\cite{Cea02} and references therein). In analytic calculations based on the
perturbation theory (PT) methods,  various resummation schemes are  used.
However, the application of different resummation techniques leads to contradictory results
about the type of the phase transition. In Ref.~\cite{Baacke:2002pi} a second order phase
transition for the $O(N)$-model was determined independently of the coupling value by
applying  some kind of resummations. The same result was also derived by using  the
renormalization group approach \cite{Nakano:2011re}. On the contrary, for the case of $O(1)$-model
the phase transition of first order was observed in the daisy, super daisy and some type
beyond resummations for the extremely weak coupling constant \cite{Bordag:2000tb}. It
was shown that various kind resummations can lack their expansion parameters near the
phase transition temperature $T \sim T_c$. The first order phase transition was also
observed within the 2PI formalism in the double-bubble approximation Ref.~\cite{Seel:2011vt}.

The discrepancies in  the analytic results can be resolved  by applying the Monte Carlo (MC) simulations on
the lattice. As numerous MC simulations showed, the second order phase transition takes place. So,
 nowadays a general belief is that the phase transition is of second order and PT
fails in this problem. However, all the performed already MC calculations  used the coupling values
$\la \ge 0.01$.  A weak-first-order phase transition determined in  Ref.~\cite{Bordag:2000tb}
for an  extremely weak couplings, has never been determined on the lattice. This notion  -
`extremely weak coupling'  - assumes some physical motivation, namely the so-called
Linde-Weinberg bound on the scalar field mass \cite{Linde:1975sw}, \cite{Weinberg:1976pe}.
For many years ago these authors have observed that in the models with negative mass
squared $m^2 \le 0$ the SSB does not happen for small values of the coupling constant
$\la \leq \la_0$ even at zero temperature. Although the actual value $\la_0$ depends on
the mass parameter entering Lagrangian, it is natural to consider small coupling values
to reach the Linde-Weinberg bound. Physically, this bound reflects an important property
of the SSB -- the existence of the range of parameters allowing the total effective potential
to be dominated by the positive radiation quantum effects instead of the negative classical
part. In this sense, the notion `Linde-Weinberg bound' is reasonable independently of
a particular model considered. In fact,  the `Linde-Weinberg bound' $\la_0$ was never
discussed within MC simulations, so the `small' values $\la \sim 0.01$ commonly used in
the simulations can occur   much greater than $\la_0$.  The goal of the present paper is to
fill the gap and to answer the question about the type of the phase transition in the
$O(1)$-model in MC simulations for values of the coupling constant much below the values
investigated already.

A well known approach to determine the type of a phase transition is to investigate the
behavior of the order parameter. For the first order transition overheated and supercooled
meta-stable states arise at the critical temperature dependently of the initial configuration.
With the initial ordered phase (cold start) the overheated configurations dominate. In
contrast, with the initial disordered phase (hot start) the supercooled configurations are
mainly observed. Since the order parameter distinguishes these alternative meta-stable phases,
collecting the statistics for different starts together one can see a hysteresis plot for the
first order transition. Of course, no hysteresis can be found for the order parameter in
case of the second order transition. These technique was successfully applied, for example,
to determine the order of the phase transition in the lattice QCD \cite{Celik:1983wz}.

In case of the $O(1)$-model, the evident order parameter is the field condensate taking non-zero
values in the phase with the broken symmetry. In MC simulations the condensate can be easily
measured as the field average over the lattice. In the present paper, we collect the statistics
for a wide interval of coupling values searching for possible hysteresis behavior. As a result,
at $\la > 10^{-3}$ no hysteresis is observed confirming the second order phase transition.
However, at $\la < 10^{-3}$ we find the hysteresis behavior, and the hysteresis
becomes more pronounced with $\la$ decreasing. Thus, we conclude that the type of the phase
transition changes at extremely weak couplings. Moreover, at $\la\sim 10^{-5}$ the ordered
phase does not occur with the hot start even at zero temperature indicating, probably, the
Linde-Weinberg  bound $\la_0$.

The paper is organized as follows. In Sect.~2 we introduce a parametrization of the $O(1)$-model
on a lattice allowing to produce stable results near the critical temperature in a wide interval
of couplings. Sect.~3 contains the results of MC simulations. Sect.~4 is devoted to conclusions.

\section{The model}

In continuous space the thermodynamical properties of the model are
described by the generating functional
\begin{equation}\label{gen-func}
Z=\int D\varphi~e^{-S[\varphi]},
\end{equation}
where $\varphi$ is real scalar field, and the action is
\begin{equation}
S =\int
dx\left(\frac12\partial_\mu\varphi(x)\partial_\mu\varphi(x)-\frac12m^2\varphi(x)^2+\frac{\lambda}{4}\varphi(x)^4\right).
\end{equation}
The standard realization of generating functional in Monte Carlo
simulations on a lattice assumes a space-time discretization and
the probing random values of fields in order to construct a
Boltzmann ensemble of field configurations. Then any macroscopic
observable can be measured by averaging the corresponding
microscopic quantity over this ensemble.

The $O(1)$-model on a lattice contains three energy scales: the mass $m$, the temperature $T$,
and the inverse lattice spacing. The mass $m$ can be chosen as a unit, then the temperature
(together with $\la$) determines the physically important values of $\varphi$, and the
lattice spacing must support successful simulation of these field values. Since the field
$\varphi$ is distributed in the infinite interval $(-\infty,\infty)$, the characteristic
scale can also move in an extremely wide interval when $\la$ changes. In principle, MC
algorithms are able to find an unknown scale in the infinite interval. However, such a
strategy requires some non-trivial parameter re-setting for different values of $\la$.
Instead, we prefer to rewrite the model in terms of dimensionless variables taking values from
finite intervals, automating also the correct choice of the lattice spacing.

First  we introduce one-to-one transformation $\varphi(U)$ to new
field variable $U(x)$ defined in the finite interval $(0,1)$. Let $U=0.5$
corresponds to $\varphi=0$ and $\varphi(U)= - ~\varphi(1-U)$. The
generating functional in terms of $U$ reads
\begin{equation}
Z=\int DU\ \det\left(\frac{\partial\varphi}{\partial
U}\right)~e^{-S[\varphi(U)]}.
\end{equation}
For Monte Carlo simulations we introduce a hypercubic lattice with
hypertorous geometry. We use an anisotropic cubic lattice
with a spatial and a temporal lattice spacing $a_s$ and
$a_t=a_s/\zeta$ with $\zeta>1$, respectively. The scalar field is
defined in the lattice sites. Transforming the Jacobian as $\det
A=\exp(\mathrm{Tr}\log A)$, the generating functional becomes
\begin{equation}
Z=\int \prod_{x} dU(x)
\exp{\left[-\left(S[\varphi(U(x))]-\sum_x\log\frac{\partial\varphi(x)}{\partial
U(x)}\right)\right]},
\end{equation}
where
\begin{eqnarray}
&&S[\varphi(U(x))]= \nonumber\\
 \sum_x\frac{a_s^4}{\zeta}&& \left[
 \left(\frac{\partial\varphi(x)}{\partial U(x)}\right)^2
 \frac{\partial_\mu U(x)\partial_\mu U(x)}{2}
 -\frac{m^2}{2}\varphi^2[U(x)]+\frac{\lambda}{4}\varphi^4[U(x)]\right].
\end{eqnarray}
The lattice forward derivative is defined as usually  by the finite
difference operation
\begin{eqnarray}
\partial_\mu U(x)\to
\frac{U(x+a_\mu\hat{\mu})-U(x)}{a_\mu},
\end{eqnarray}
where $a_\mu$ is the lattice spacing in the $\mu$ direction,
$\hat{\mu}$ is the unit vector in the direction indicated by
$\mu$.

In the case of pure condensate field the action is determined by the
potential
\begin{eqnarray}
\tilde{V}[U]=\left[-\log\frac{\partial\varphi}{\partial U}+
\frac{a_s^4}{\zeta}
\left(-\frac12m^2\varphi^2[U]+\frac{\lambda}{4}\varphi^4[U]\right)\right].
\end{eqnarray}
This potential is  topologically equivalent to the potential
$V(\varphi)=-m^2\varphi^2/2$ $+\lambda\varphi^4/4$. It has one
local maximum at $U=0.5$ and two symmetric global minima at $U_0$
and $1-U_0$. The spread between the values of the potential at
local maximum and global minima is
\begin{equation}\label{DV}
\Delta_V= \log\frac{\partial\varphi/\partial
U|_{U=U_0}}{\partial\varphi/\partial U|_{U=0.5}}
-\frac{a_s^4}{\zeta}
\left(-\frac12m^2\varphi^2[U_0]+\frac{\lambda}{4}\varphi^4[U_0]\right).
\end{equation}

The quantities $U_0$ and $\Delta_V$ play a crucial role in
Monte Carlo simulations.

Considering the phase transition, one must guarantee that the
Monte Carlo algorithm meets the field values compatible with both
the phases to choose. If $U_0\to 0.5$, then the broken phase can
be missed since the corresponding field values are extremely rare
events. On the other hand, $U_0\to 0$ ($U_0\to 1$) washes the
unbroken phase out.
So, to study the phase transition in the model, we choose the
following conditions:
\begin{equation}\label{cond-m}
U_0=0.25, \qquad \Delta_V=1.
\end{equation}
Thus, the half of generated field values will be between the
global minima of the `effective' potential, and no phase will be
accidentally missed. The probability to prefer condensate or
non-condensate values will be of order $\sim 0.5$ ensuring the
fast convergence of Monte Carlo algorithm. As it will be shown,
these conditions successfully works for $10^{-5}<\lambda<0.5$. For
the larger (smaller) $\lambda$ the condensate becomes too weak
(strong) and the numeric values in (\ref{cond-m}) must be
reconsidered.

To satisfy two conditions (\ref{cond-m}) we use a convenient
two-parameter function
\begin{eqnarray}\label{funct}
\varphi[U]&=&m\xi\mathrm{arctanh}\left[\eta(2U-1)+(1-\eta)(2U-1)^3\right]
\end{eqnarray}
with $\xi>0$ and $\eta>0$. The values of $\xi$ and $\eta$ have to
be found as the solution of equations $d\tilde{V}/dU|_{U=U_0}=0$
and (\ref{DV}). These equations can be written as
\begin{eqnarray}\label{syst-1}
&&\frac{2{\cal F}[U_0]-{\cal G}[U_0]}{\left({\cal F}[U_0]-{\cal G}[U_0]\right)^2}=z =\frac{\lambda\zeta}{m^4 a_s^4},\\
&&\frac{\lambda}{m^2 z}\varphi^2[U_0] ={\cal F}[U_0]-{\cal
G}[U_0],\label{syst-2}
\end{eqnarray}
where $z$ is a dimensionless parameter of the model,
\begin{eqnarray}\label{auxF}
{\cal
K}[U]&=&\frac{1-\eta(2U-1)-(1-\eta)(2U-1)^3}{1+\eta(2U-1)+(1-\eta)(2U-1)^3},\\
{\cal
F}[U]&=&
\left(\frac{{\cal K}''[U]{\cal K}[U]}{({\cal K}'[U])^2}-1\right)\log{\cal K}[U],\\
{\cal G}[U]&=& 4\left(\log\frac{-{\cal K}'[U]}{4\eta{\cal
K}[U]}-\Delta_V\right),
\end{eqnarray}
where the primes denote derivatives. Eq.~(\ref{syst-1}) gives
$\eta(z)$, then $\xi$ can be found from (\ref{syst-2}). There is
no physical solution for $z<z_\mathrm{min}$. This forbidden
interval corresponds to low temperatures which cannot be reached
within the chosen parametrization. Finally the lattice action is
\begin{eqnarray}
{S}[U(x)]&=&\sum_x\sum\limits_{\mu} \left[
Y\sqrt{\frac{z}{\zeta\lambda}}
 \left(\frac{{\cal K}'[U(x)]}{{\cal K}[U(x)]}\right)^2
 \left(\frac{U(x+a_\mu\hat{\mu})-U(x)}{a_\mu/a_s}\right)^2\right]
 \nonumber\\
&+&
 \sum_x\left[-\frac{1}{4}{\cal G}[U(x)]  -Y\log^2 {\cal K}[U(x)]
 +Y^2 z\log^4 {\cal K}[U(x)] +V_0
\right],\nonumber\\
Y&=&\frac{{\cal F}[U_0]-{\cal G}[U_0]}{2\log^2{\cal K}[U_0]},\\
V_0&=& -\Delta_V -\log(2m\xi\eta).
\end{eqnarray}
The constant part of the action $V_0$ is completely unimportant
for calculations and can be omitted, since Monte Carlo algorithm
is based on the difference between the actions of modified and
initial field configurations.

\begin{figure}
    \centering
    \includegraphics[bb=211 226 548 440,width=0.6\textwidth]{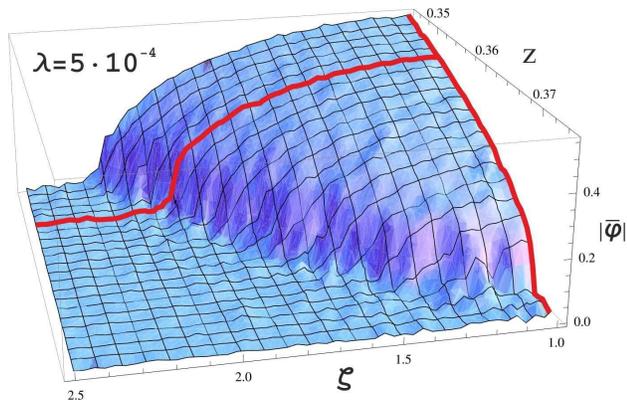}
  \caption{$|\bar\varphi|$ in the units of classical condensate
  $m/\sqrt{\lambda}$ for $\lambda=5\cdot 10^{-4}$ on lattice $16^4$}
  \label{fig:lambdaZzeta}
\end{figure}

By varying $\zeta$ it is possible to change $a_t$, while keeping
$a_s$ fixed. Consequently the temperature $T\sim \zeta$ can be
changed continuously at fixed $a_s$. The field condensate,
$\bar{\varphi}$, is measured as the average of $\varphi(x)$ over
the lattice. In Fig.~\ref{fig:lambdaZzeta} we plot $|\bar\varphi|$ in
the units of classical condensate $m/\sqrt{\lambda}$ for
$\lambda=5\cdot 10^{-4}$ and $16^4$ lattice. Lower values of
$\zeta$ and $z$ corresponds to lower temperatures. One can see
evident phase transition with the field condensate growing with
the decreasing temperature. Since the clear positive or negative
values of $\bar\varphi$ appear in lattice configurations in the
broken phase, we conclude about the absence of domains and will
use $|\bar\varphi|$ in plots.

To determine the type of the phase transition we consider in
details three slices of two-dimensional function $|\bar\varphi|(z,\zeta)$ at
fixed $z=0.35$, $z=0.5$ (for $\lambda=5\cdot 10^{-5}$), and $\zeta=1$. For example, the slices
are shown in Fig.~\ref{fig:lambdaZzeta} as bold red lines ($\lambda=5\cdot 10^{-4}$).
We compute the field condensate with the hot and cold starts for different
$\zeta$. A hysteresis behavior means a first order phase transition.

\section{The Monte Carlo simulation results}
To estimate the order of the phase transition a large amount of simulation
data must be prepared. The simulation requires a fairly powerful
computing resources, especially for large lattices. To speed up essentially the generation of data
we apply a GPU cluster as a computational platform. It consists
of ATI Radeon GPUs: HD6970, HD5870, HD5850 and HD4870 with the peak performance up
to 11 Tflops. The low-level AMD Intermediate Language (AMD IL) is used in order
to obtain the maximal performance of used hardware.
A trivial parallelization scheme is implemented for cluster computation.
Some technical details of MC simulations on the ATI GPUs and a review
of the AMD Stream SDK are given in Ref.~\cite{Demchik:2009ni} and
references therein.

In the MC simulations, we use lattices
of different sizes up to $64^4$. Most statistics are obtained for lattices
$32^4$ and $16^4$, the qualitative behavior are checked on larger lattices.
\textsf{RANLUX} pseudo-random number generator is used in
the MC kernel, all the key results are checked with \textsf{RANMAR}
generator \cite{Demchik:2010fd}. Lattice data are stored with a
single precision. MC updating are also performed with the single
precision whereas all averaging measurements are performed with
the double precision to avoid the accumulation of errors.

The system passed 5000 MC iterations for
every run to be thermalized, then we used 1024 MC configurations (separated
by 10 updates) for measuring.

\begin{center}
\begin{figure}[tb]
    \centering
    \includegraphics[bb=49 121 704 487,width=0.4\textwidth]{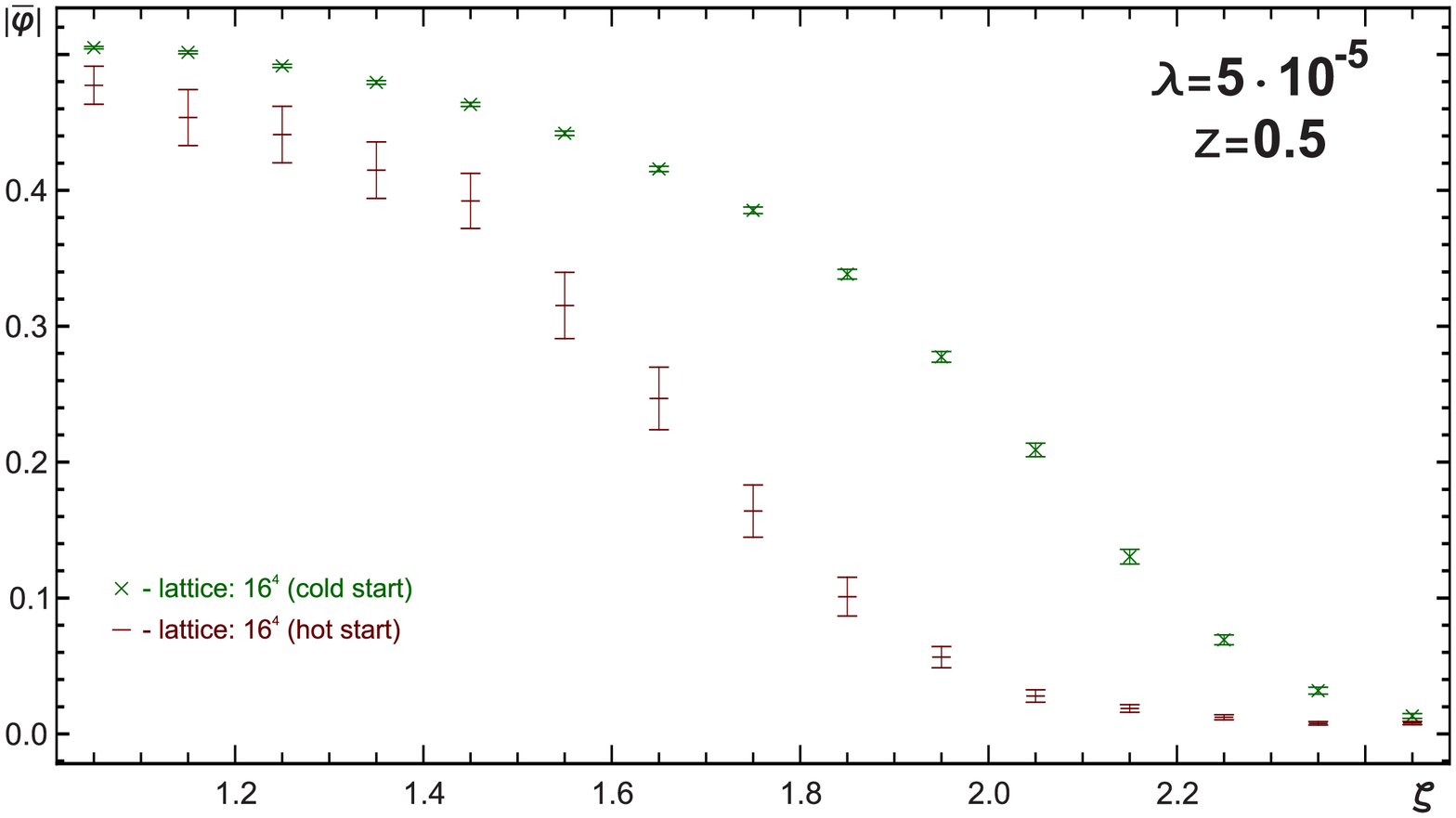}
\hskip 1cm
    \includegraphics[bb=49 121 697 487,width=0.4\textwidth]{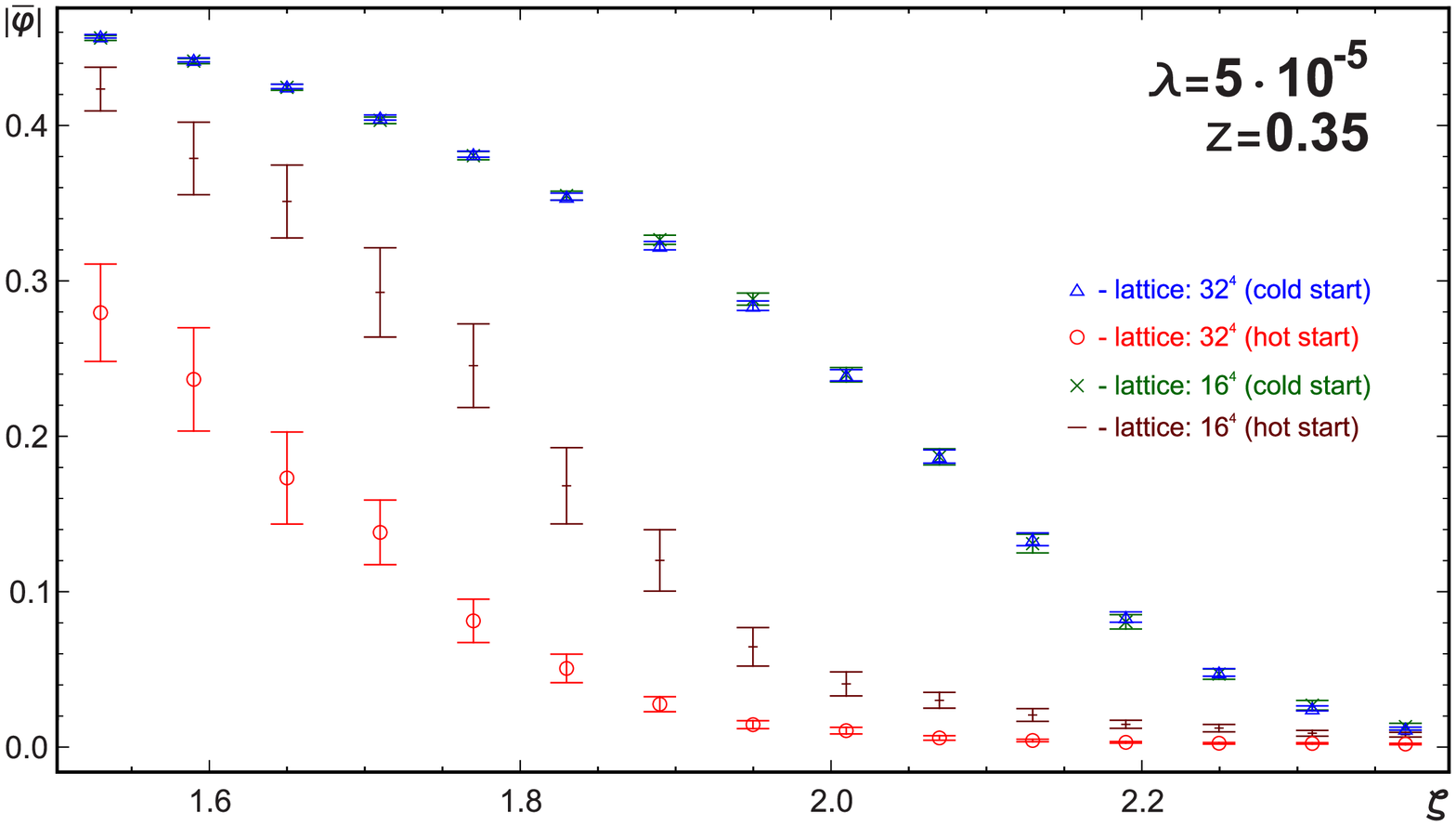}
\\\vskip 0.5cm
    \includegraphics[bb=59 138 708 504,width=0.4\textwidth]{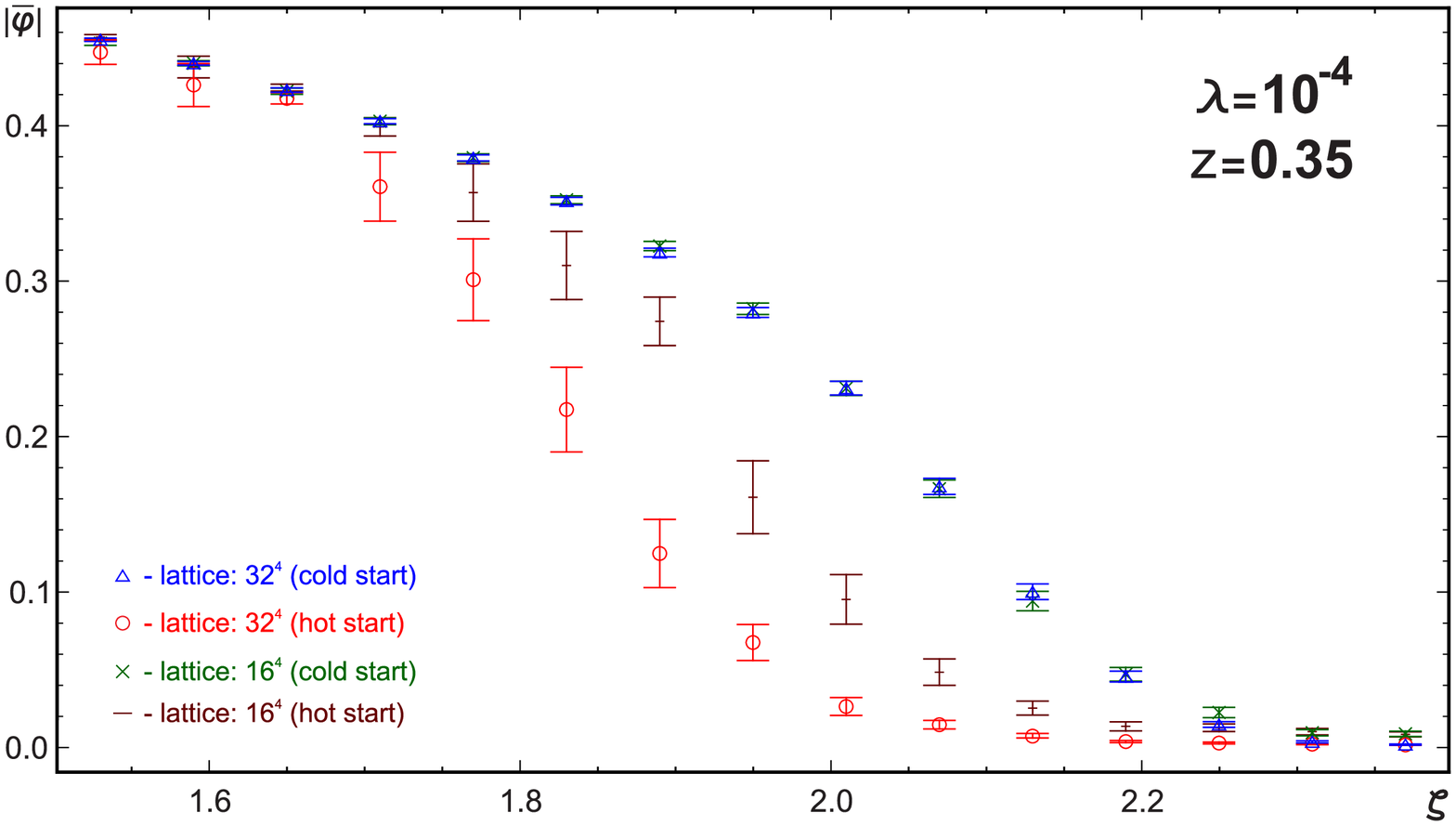}
\hskip 1cm
    \includegraphics[bb=59 164 708 504,width=0.4\textwidth]{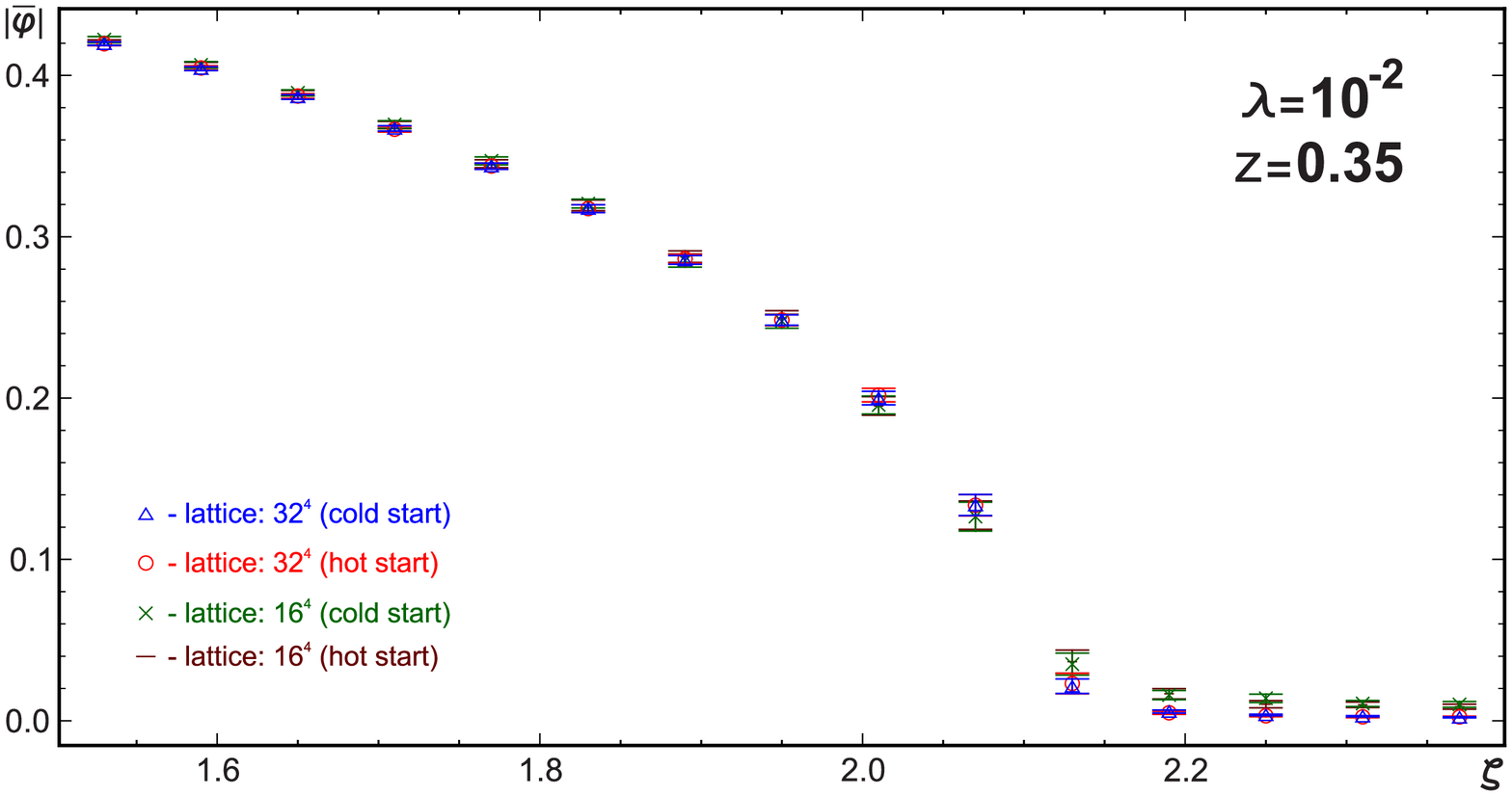}
  \caption{Temperature dependence of the field condensate $|\bar\varphi|$ for the lattices
$16^4$ and $32^4$ at $z=0.35$ and $z=0.5$ for $\zeta=[1.5; 2.5]$.}
\label{fig:lambdas}
\end{figure}
\end{center}

The temperature dependence of $|\bar\varphi|$ for the lattices
$32^4$ and $16^4$ at $z=0.35$ and $z=0.5$ for $\zeta=[1.5; 2.5]$ is present in
Fig.~\ref{fig:lambdas}. The whole data set for every plot is divided
into 15 bins. Different initial conditions are shown as:
\begin{itemize}
  \item \textsf{hot start} -- red circles (lattice $32^4$) and brown dashes (lattice $16^4$);
  \item \textsf{cold start} -- blue triangles (lattice $32^4$) and green crosses (lattice $16^4$).
\end{itemize}
The mean values and 95\% confidence intervals are presented by corresponding pointers for each bin. Every
bin contains 150 points.

As it is seen from Fig.~\ref{fig:lambdas}, for $\lambda=0.01$
the temperature dependence of field condensate is not sensitive to
the initial configuration. Both cold and hot starts lead to the same
behavior of field condensate for various $\zeta$. This meas a second order phase transition.

For smaller values of  $\lambda$ the overheated configurations
occur in the broken phase for the hot start. That is, different
starts demonstrate a hysteresis behavior. This corresponds to a
phase transition of the first order.

With further decreasing of $\la$, for $\la \leq \la_0 \sim 10^{-5}$ the behavior
of cold and hot starts is completely separated and independent of the temperature.
This is plotted in Fig.~\ref{fig:lambdas}. Such type property probably means that the SSB
does not happen even at zero temperature and corresponding value of $\la_0$
can be identified with the Linde-Weinberg low bound.

\begin{figure}
    \centering
    \includegraphics[bb=49 149 704 487,width=0.6\textwidth]{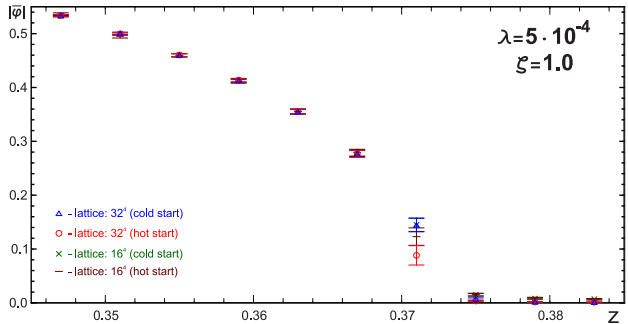}
  \caption{$|\bar\varphi|$ in the units of classical condensate
  $m/\sqrt{\lambda}$ for $\lambda=5\cdot 10^{-4}$ and $\zeta=1$ on lattices $16^4$ and  $32^4$}
  \label{fig:lambdaZ}
\end{figure}

As it is seen in Fig.~\ref{fig:lambdas}, two different slices $z=0.35$ and $z=0.5$ demonstrate the same behavior.
The calculations for the slice $\zeta = 1$ reproduce the described above results again (see Fig.~\ref{fig:lambdaZ}).
Thus, the hysteresis occurs at small $\la$ independently of the model parameters $z$ and $\zeta$.

\section{Conclusion}

As it was discovered in the MC simulations, the temperature phase transition in $O(1)$ $ \phi^4$ model is strongly dependent on the coupling value
$\la$. There is a low bound $\la_0 = 10^{- 5}$ determining the range where SSB is not realized. Close to this value in the interval $10^{-5} \leq \la \leq 10^{- 3}$ the phase transition is  first order. For larger values of $\la$ the second order phase transition happens. These types of the behavior  have been determined on the lattices of different sizes independently of the internal model parameters. Our calculation procedure was developed to accelerate the MC procedure in the domain of parameters close to the transition for a wide range of coupling. For usually considered values of $\la \sim 0.01 - 0.1$ it gives the results coinciding with the ones existing in the literature and signalling the second order phase transition. To our knowledge, systematic investigations for smaller values of coupling have not been carried out yet.

 Our observations, in particular, may serve as a guide for the applicability of different kind resummations in perturbation theory. In fact, we see that the daisy and super daisy resummations give qualitatively correct results for small values of $\la$. For larger values they become non-adequate to the second order nature of the phase transition. In this case other more complicated resummation schemes should be used.

 The change of the phase transition type dependently of the coupling value is not a new phenomenon. For instance, in the standard model of elementary particles it is well known that the electroweak phase transition is of  first order for small $\la$ and it converts into a cross-over or even  second order one  for sufficiently large values of $\la$. We have observed that this happens  even in the simple model with one coupling.

In the present investigation, we  concentrated mainly on the qualitative aspects of converting the phase transition type due to the change of the coupling values. So, we skip an ubiquitous procedure relating the lattice and physical variables, as unessential.


\noindent{\bf Acknowledgements.} The authors are grateful to P.M.Stevenson for useful suggestions.
One of us (VD) was supported by DFG under Grant No BO1112/17-1. He also thanks the Institute for
Theoretical Physics of Leipzig University for kind hospitality.

\end{document}